\def\draftversion{false}

\RequirePackage{ifthen}
\ifthenelse{\equal{\draftversion}{true}}{
	\documentclass[aps,prb,10pt,galley,amsmath,amssymb,
	floatfix,
	citeautoscript]{revtex4}
}{
	\documentclass[aps,prb,10pt,twocolumn,amsmath,amssymb,
	citeautoscript]{revtex4-1}
}

\usepackage{graphicx}
\usepackage[usenames,dvipsnames]{color} 
\usepackage{bm} 
\usepackage[update,prepend]{epstopdf}
\usepackage{soul} 
\usepackage{dcolumn} 

\def\Green#1{\textcolor{OliveGreen}{#1}}

\def\comment#1{}

\ifthenelse{\equal{\draftversion}{true}}{
	\marginparwidth 2.7in
	\marginparsep 0.5in
	\newcounter{comm} 
	\def\commnext{\stepcounter{comm}}
	\def\commtext{{\bf\color{blue}[\arabic{comm}]}}
	\def\commmar{{\bf\color{blue}[\arabic{comm}]}}
	\def\dvm#1{\commnext\marginpar{\small DV\commmar: #1}\commtext}
	\def\jkm#1{\commnext\marginpar{\small JK\commmar: #1}\commtext}
	\def\krm#1{\commnext\marginpar{\small KR\commmar: #1}\commtext}
	\def\mlab#1{\marginpar{\small\bf #1}}
	
}{
	\def\dvm#1{}
	\def\jkm#1{}
	\def\krm#1{}
	\def\mlab#1{}
	
}

\begin{document}

\title{Negative piezoelectric response of van der Waals layered bismuth tellurohalides}

\author{Jinwoong Kim}
\author{Karin M. Rabe}
\author{David Vanderbilt}
\affiliation{
  Department of Physics and Astronomy,
  Rutgers University, 
  Piscataway, New Jersey 08854-8019, USA
}

\date{\today}
\begin{abstract}
The polarization and piezoelectric response of the BiTe$X$ ($X$=Cl,
Br, and I) layered tellurohalides are computed from first principles.
The results confirm a mixed ionic-covalent character of the bonding,
and demonstrate that the internal structure within each triple layer
is only weakly affected by the external stress, while the changes in
the charge distribution with stress produce a substantial negative
piezoelectric response.  This suggests a new mechanism for negative
piezoelectric response that should remain robust  even in ultra-thin film
form in this class of materials.
\end{abstract}

\maketitle

\section{Introduction}
\label{sec:intro}

Conventional soft-mode ferroelectric materials exhibit a structural
transition from a paraelectric to a polar phase as the temperature
falls through a critical $T_{\rm c}$, below which a polar phonon
mode freezes in to generate a ferroelectric ground
state.\cite{Haertling,Resta1994,Dawber2005}  
In general, this polar phonon mode stiffens as the lattice constants are reduced, 
and consequently the magnitude of the polarization decreases with
compressive strain. 
This corresponds to a positive piezoelectric
response, $d_{33} > 0$, where by convention we consider the polar
variant with $P>0$.
Recently, however, Liu and Cohen proposed a different mechanism by which
materials can exhibit a negative piezoelectric response, and identified
hexagonal $ABC$ ferroelectrics as a class of materials in which the
lattice contribution to the piezoelectricity is positive, but the
frozen-ion or electronic contribution is negative and
larger.\cite{Liu2017} Negative piezoelectricity has
also been experimentally demonstrated in ferroelectric
polymers\cite{Katsouras2015} and in organic molecular
ferroelectric materials.\cite{Urbanaviciute}

\begin{figure}[b]
        \centering
	\includegraphics[width=8.0cm]{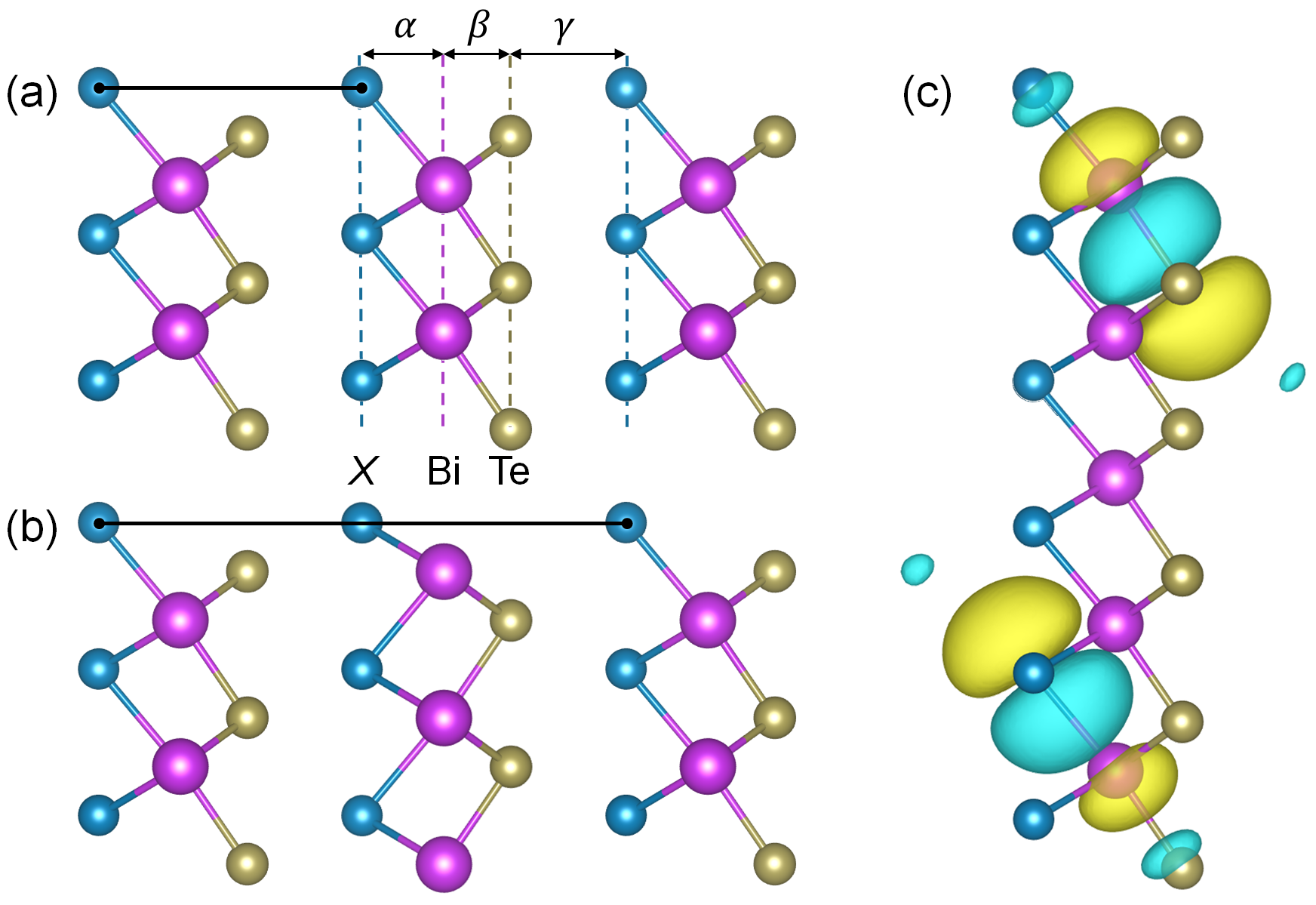}
	\caption{Layered structure of BiTe$X$ for (a) $X$\,=\,I
	or Br in the $P3m1$ structure, and (b) $X$\,=\,Cl in
	the doubled-cell $P6_3mc$
	structure. Horizontal solid lines indicate $c$ 
	lattice	constants. Distances $\alpha$ and $\beta$ denote
	Bi-$X$ and Bi-Te intralayer distances, while $\gamma$ is
	the separation between triple layers. (c) Wannier functions
	constructed from $p$ bands,
	having Bi-$X$ (lower) and Bi-Te (upper) bond-orbital character.}
	\label{fig:struct}
\end{figure}

BiTe$X$ ($X$\,=\,Cl, Br, and I) compounds have attracted
considerable recent interest as strongly polar
quasi-2D materials.
The breaking of inversion symmetry results from the layer geometry,
in which a central Bi layer is neighbored by a Te layer on one side
and a halide layer on the other side, forming a triple layer (TL)  
as shown in Fig.~\ref{fig:struct}. 
The TLs are bonded to each other by weak van der 
Waals interactions, implying easy exfoliation and a soft mechanical 
response under uniaxial stress.  
The TLs are stacked so that the Te layer is always on the same side of 
Bi, with one-TL periodicity resulting in a polar $P3m1$ space group
for $X$\,=\,Br and I, and two-TL periodicity resulting in a 
polar $P6_3mc$ 
space group for $X$\,=\,Cl.
Because of the strongly broken
inversion symmetry combined with strong spin-orbit coupling, these
materials are of interest for their large
bulk Rashba effect, with potential spintronic applications.
\cite{Ishizaka2011,Bahramy2011,Monserrat2017}
BiTeI has also been much discussed for its topological properties,
since it has been predicted to undergo a topological phase transition to a
strong topological-insulator phase under pressure, mediated by
a narrow but topologically robust Weyl semimetal
phase\cite{Murakami2007,Bahramy2012,Liu2014,Murakami2017},
and to exhibit an enhanced nonlinear Hall conductivity.\cite{Facio2018}

The broken centrosymmetry of the crystal also naturally suggests the 
possibility of ferroelectricity or piezoelectricity.  The polarization is 
not associated with a polar distortion of a nearby high-symmetry 
reference structure, and is inherently not switchable, since the 
bonding within the TL is much too
strong to allow a structural reversal under applied electric field.  
However, as these systems are mechanically soft, there is a 
marked change in structure under applied stress, which can be
expected to result in a change in polarization and corresponding 
piezoelectric response.

In this study, we investigate the electric polarization and
piezoelectric response of BiTe$X$ by using first-principles
calculations. We compare the calculated dipole moments with two
plausible models that anticipate opposite directions of the dipole
moment, deciding in favor of the one that treats the BiTe unit as
more covalently than ionically bonded. We will see
that while structurally, the BiTe$X$ TLs behave as relatively rigid 
units, internal charge rearrangement under applied uniaxial strain or
stress leads to a substantial negative piezoelectric response. 
This suggests a new mechanism of piezoelectricity that may be
widely applicable to a broad class of insulating materials based on
layered van der Waals stacking of polar constituents.

\section{Methods}
\label{sec:methods}

The polarization and piezoelectric response
of BiTe$X$ are determined from 
first-principles calculations carried out using the \small{VASP}
package.\cite{Kresse96a,Kresse96b} 
The pseudopotentials are of the projector-augmented-wave type as 
implemented in \small{VASP},\cite{Blochl94,KressePAW}
with valence configurations
6$s^2$6$p^3$ for Bi,
5$s^2$5$p^4$ for Te,
and 3$s^2$3$p^5$, 4$s^2$4$p^5$, and 5$s^2$5$p^5$
for Cl, Br, and I, respectively.
The exchange-correlation functional 
is described by the modified Perdew-Burke-Ernzerhof generalized gradient 
approximation for solids (PBEsol). \cite{PBEsol}
The plane-wave cut-off energy is set to $400$\,eV.
The Brillouin zone sampling grid is
12\,$\times$\,12\,$\times$8
for the 1-TL periodic $P3m1$ structure and
12\,$\times$\,12\,$\times$4
for the 2-TL periodic $P6_3mc$ structure;
relative energy differences between the two structures were
obtained by computing both in the doubled-cell structure
with the 12\,$\times$\,12\,$\times$4 grid.
Spin-orbit coupling is included in all calculations.
The structural coordinates are relaxed 
within a force threshold of 1.5\,meV/\AA.
The electric polarization is computed 
using the Berry-phase method,
\cite{King-Smith,Vanderbilt} and the Wannier charge 
centers\cite{Marzari2012} are obtained
using the \small{VASP-Wannier90} interface.\cite{Mostofi}
The maximal localization of the Wannier functions is carried
out separately for the $s$ and $p$ bands to avoid $sp_3$ hybridization.


For a crystal composed of weakly coupled molecules or layers, 
it is natural to compute the polarization from the dipole moment of the 
individual unit. 
For a periodic system, this value can be quantitatively obtained by computing
the Berry phase polarization ~\cite{King-Smith,Vanderbilt,Resta}, 
where the branch choice arising from the quantum of polarization can
resolved by choosing the value closest to that estimated by the 
dipole moment integral, or by using the Wannier center formulation and
choosing the Wannier centers to be within the individual unit.
In this work, we consider only $p=p_3$, the dipole moment per the unit cell measured
along the stacking direction $\hat{e}_3$, and adopt the convention that
the polarization $P_3$,
electric field $\mathcal{E}_3$,
strains $\eta_3 = c/c_0 -1$, and stresses $\sigma_3$ (Voigt notation for
$\eta$ and $\sigma$), will also 
be written without the subscript for simplicity.

We calculate various piezoelectric responses following the standard
definitions.~\cite{Ballato,Nye2012,Wu2005}
The piezoelectric stress tensor 
elements $e_{\alpha j}$ are defined in terms of
the derivative of stress with respect to the electric field,
or equivalently,
polarization with respect to strain,
\begin{equation}
e_{\alpha j}
= \left. -\dfrac{\partial \sigma_j}{\partial \mathcal{E}_{\alpha}} \right\vert_{\eta}
= \left. \dfrac{\partial P_{\alpha}}{\partial \eta_j} \right\vert_{\mathcal{E}},
\end{equation}
while the piezoelectric strain tensor
elements $d_{\alpha j}$ are related to the
derivative of strain with electric field,
or equivalently, the derivative of polarization with respect to
stress,
\begin{equation}
d_{\alpha j} 
= \left. -\dfrac{\partial \eta_j}{\partial \mathcal{E}_{\alpha}} \right\vert_{\sigma}
= \left. \dfrac{\partial P_{\alpha}}{\partial \sigma_j} \right\vert_{\mathcal{E}}.
\end{equation}
To calculate the piezoelectric stress response, 
the polarization $P$ is calculated on a grid of strains $\eta$
with the in-plane lattice constant fixed to the zero-stress value,
and is fitted to a polynomial to obtain the derivative corresponding to the
piezoelectric response $e_{33}$.  
For the same grid of strains $\eta$, we compute the optimized value
of $a$ at each $\eta$, and then using the values of stress
and polarization reported by VASP at each $\eta$, we
fit the results in order to extract the value of
the piezoelectric $d_{33}$ coefficient.
In addition, we compute a mixed response
$d_{33}^{\rm epi}$ by carrying out a similar fitting procedure but at fixed
in-plane lattice constant.

\section{Results}

\subsection{Structure and polarization}
\label{sec:struct}

BiTeI and BiTeBr crystallize in the hexagonal structure illustrated
in Fig.~\ref{fig:struct}(a), space group $P3m1$ (\#156), with
three atoms per cell.  BiTeCl has the same internal layer structure,
but alternate TLs are rotated 180$^\circ$ about $\hat{e}_3$ on an
axis passing through the $X$ atom, as shown 
in Fig.~\ref{fig:struct}(b),
resulting in a doubled six-atom unit
cell belonging to space group $P6_{3}mc$ (\#186).

Our computed structural parameters for these three materials,
together with the Berry-phase polarization $P$,
are given in Table~\ref{tab:1polar}. 
In the case of $X$\,=\,Cl, we carried out calculations in both
the $P3m1$ and $P6_{3}mc$ structures; the former is designated
with a prime (BiTeCl$'$) as a reminder that it is not the experimental
ground-state structure.
We see rather obvious trends in that the volume and the Bi-$X$
distance $\alpha$ shrink as $X$ becomes more electronegative, while
the Bi-Te distance $\beta$ remains roughly constant.
The trend in going to $X$\,=\,Cl is most consistent
when the same structure is assumed (first three rows of the table).
The change to the doubled-cell $P6_{3}mc$ in the last row is
generally small, showing that the stacking sequence does not have
a strong effect on the structural parameters.

The calculated $c$ lattice constant of BiTeI is close to 
the experimental value of 6.854\,\AA.\cite{SHEVELKOV1995}
We tested several different exchange-correlation potentials
including some with van der Waals corrections, but we find
that our use of PBEsol produces the closest agreement
for the $c$ lattice constant,
with an error of 0.5\%, compared to the other ones we tested
(5.9\% for PBE\cite{PBE}, 
3.7\% for PBE+TS\cite{Tkatchenko},
and 2.2\% for SCAN\cite{SCAN}).
This is consistent with a previous theoretical report\cite{Monserrat2017}
in which the PBEsol functional was found to give the most accurate
prediction of the BiTe$X$ structure.

\begin{table}
\caption{Calculated structural parameters of BiTe$X$ ($X$=I, Br, Cl).
$V$ is cell volume, $a$ and $c$ are in-plane and out-of-plane lattice
constants, $\alpha$ and $\beta$ are Bi-$X$ and Bi-Te layer spacings,
and $P$ is polarization. The prime on BiTeCl$'$ denotes the results
for the same $P3m1$ structure as for $X$\,=\,I and Br; the unprimed
version is for the ground-state $P6_{3}mc$ structure.\\
\label{tab:1polar}}
\begin{ruledtabular}
\begin{tabular}{lcccccc}
& $V$ (\AA$^3$) & $a$ (\AA) & $c$ (\AA) & $\alpha$ (\AA) & $\beta$ (\AA) &
   $P$ (C/m$^2$) \\
\hline
BiTeI    & 111.5 & 4.343 &\, 6.823 & 2.104 & 1.721 & 0.069 \\
BiTeBr   & 102.6 & 4.270 &\, 6.499 & 1.871 & 1.754 & 0.100 \\
BiTeCl$'$&\, 97.5& 4.235 &\, 6.275 & 1.677 & 1.767 & 0.107 \\  
BiTeCl   & 195.5 & 4.239 & 12.563  & 1.667 & 1.765 & 0.099 
\end{tabular}
\end{ruledtabular}
\end{table}

We computed the ground-state energies for each of the
materials in both the $P3m1$ and $P6_3mc$ structures, finding that the
doubled-cell $P6_3mc$ structure
is higher in energy by 7.3, 4.8, and 5.1\,meV for
BiTeI, BiTeBr, and BiTeCl, respectively.
This correctly predicts the 1-TL ground state structure for BiTeI
and BiTeBr, but it does not account for the observation of the
2-TL structure of BiTeCl.  However, the energy differences are
small, and are near the limit of our first-principles resolution.
We speculate that it may be necessary to take differences
in vibrational entropy into account in order to explain the
observed structure of BiTeCl.
In any case, as noted above, a comparison of the BiTeCl and BiTeCl$'$
results in Table~\ref{tab:1polar} shows that
the structural properties are not
very sensitive to the choice of space-group structure, and we report
results for BiTeCl in both structures.

Figure~\ref{fig:struct}(c) shows two of the maximally localized
Wannier functions constructed from the $p$ bands of BiTeI,
rendered using the \small{VESTA} software package~\cite{VESTA}.  We
see a somewhat asymmetric bond orbital composed of Bi and $X$ $p$
orbitals at bottom, and a somewhat more symmetric bond orbital made of
Bi and Te $p$ orbitals at top.  Both show significant covalent
bonding, but the greater asymmetry of the Bi-$X$ bond orbital is
consistent with a stronger ionic character, as expected from the
stronger electronegativity of the halide $X$ atom.
The trend in the strength of covalency is also evident from the values of
$\alpha$ and $\beta$ reported in Table~\ref{tab:1polar}.
Not surprisingly, the $\beta$ value (Bi-Te spacing)
remains roughly constant, while the $\alpha$ value
(Bi-$X$ spacing) shrinks significantly in going from $X$\,=\,I
to Br to Cl, with the increasing electronegativity of the $X$ ion.

The results for the electric polarization for each of the
three materials, computed using the Berry-phase approach as described in the Methods
section, are presented in the last column of
Table~\ref{tab:1polar}.
Not surprisingly, the polarization, being a dipole moment per unit volume,
increases as the volume decreases, but this does not account for
all of the variation.
The interpretation of the sign and magnitude of the polarization
is the topic of the next subsection.

\subsection{Interpretation of the polarization}
\label{sec:interp}

\begin{figure}
        \centering
	\includegraphics[width=8.0cm]{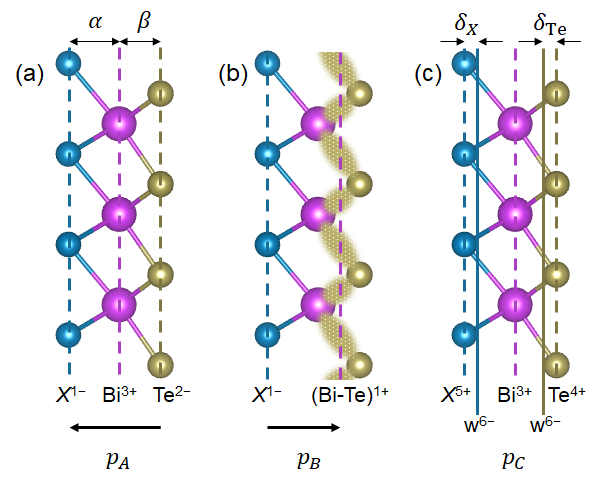}
	\caption{(a-b) Schematic view of two simple models of the
	 polarization in BiTe$X$. 
	(a) Fully ionic model suggests a negative dipole moment.
    (b) Model with a covalent Bi-Te unit suggests a positive
        dipole moment.  
	(c) Wannier function perspective. Solid vertical
	lines indicate Wannier center locations,
	with $\delta _{X,\textrm{Te}}$ denoting their
	displacements relative to the neighboring anion
	layers (dotted vertical lines).}
	\label{fig:models}
\end{figure}

The computed polarization reported in Table~\ref{tab:1polar} is positive,
so the dipole moment of the layers points along the direction 
from the halide to the Te. 
%
Surprisingly, two simple models, shown in Fig.~\ref{fig:models}(a) and (b),
predict opposite signs of the polarization.
%
Model A, the fully ionic model shown in Fig.~\ref{fig:models}(a),
assumes that
the ions keep their nominal valence, in which case the dipole moment is 
estimated as
\begin{equation}
p_{\rm A}=e(\alpha-2\beta).
\label{eq:ModelA}
\end{equation}
(recall that $\alpha$ and $\beta$ are the Bi-$X$ and Bi-Te layer spacings
respectively).  From the figure, it is clear that the dipole moment
would then point to the left (negative, in our convention) because
of the excessive negative charge of Te$^{-2}$ compared to $X^{-}$.

Model B, shown in Fig.~\ref{fig:models}(b),
assumes a strong covalent bond between neighboring Bi and Te layers,
and treats this pair of layers as a single unit with an overall
valence of $+1$.
The dashed vertical line at right in Fig.~\ref{fig:models}(b)
indicates an average position of this Bi-Te unit, taken to be
midway between the Bi and Te planes.
In this model the dipole is predicted to be
\begin{equation}
p_{\rm B}=e(\alpha+\beta/2)
\label{eq:ModelB}
\end{equation}
which is clearly positive, in contrast with the prediction of the
previous model.


\begin{table}[t]
	\caption{Dipole moment of the triple atomic layer in bulk BiTe$X$,
        in units of $e${\AA}, for the three models
	in Fig.~\ref{fig:models}. Last two columns give the values of
        $\delta_X$ and $\delta_{\rm Te}$, describing the displacements of
        the $p$-band Wannier centers from nearby ionic planes,
        in units of {\AA}.
	BiTeCl$'$ refers to the same $P3m1$ structure as for $X$\,=\,I
	and Br, although it is not the observed ground state for $X$\,=\,Cl.}
	\label{tab:2model}
	\begin{ruledtabular}
	\begin{tabular}{ l c c c c c c } 
		& $p_{\rm DFT}$ & $p_{\rm A}$ & $p_{\rm B}$ & $p_{\rm c}$
                & $\delta_X$  & $\delta_{\rm Te}$\\ [0.5ex] 
		\hline
		BiTeI    & 0.48 & $-$1.34 & 2.96 & 0.43 & 0.24 & 0.54 \\
		BiTeBr   & 0.64 & $-$1.64 & 2.75 & 0.56 & 0.19 & 0.55 \\
		BiTeCl$'$& 0.61 & $-$1.86 & 2.56 & 0.57 & 0.15 & 0.55 \\
		BiTeCl   & 0.61 & $-$1.86 & 2.55 & 0.53 & 0.15 & 0.55
	\end{tabular}
	\end{ruledtabular}
\end{table}

Table~\ref{tab:2model} reports the values of the dipole moment as computed
from first principles ($p_{\rm DFT}$), as well as the values computed
from Models A and B. Neither of these models
gives a value for the dipole that is close to the first-principles value.
The first-principles result is not far from an average of the two,
suggesting that Models A and B can be taken as describing
two end points corresponding to extreme ionic and mixed ionic-covalent
bonding states, respectively.

As described
in Sec.~\ref{sec:methods}, the polarization is given exactly
in terms of the coordinates of the ions and
those of the Wannier centers of the occupied bands.  The positions of
the Wannier centers constructed from the occupied $p$
bands, shown earlier in Fig.~\ref{fig:struct}(c), are indicated by
the solid vertical lines in  Fig.~\ref{fig:models}(c).
This approach would be exact if the information on the Wannier
centers constructed from the occupied $s$ bands were included as
well, but we assume these to coincide with the atomic
coordinates;
this is a reasonable
approximation since the $s$ bands are well separated and weakly
hybridized with other bands.
There are thus two
additional parameters taken from first principles, namely the shifts
$\delta_X$ and $\delta_{\rm Te}$, relative to the anion coordinates,
of the Wannier centers constructed from the occupied $p$ bands.
 We then include
the $s$ charge $-2e$ into our definition of the core charges, which
become +4, +3, and +5 for Te, Bi, and $X$, respectively.
The remaining twelve electrons form anion $p$-like Wannier functions
whose Wannier-center positions are illustrated in
Fig.~\ref{fig:models}(c). Six of these are associated
with Wannier centers displaced by $\delta_{\textrm{Te}}$ from the
Te centers, and the other six are centered a distance $\delta_X$ from the
$X$ centers, measured along the $z$ direction.  Accounting for
all of the ionic and Wannier contributions shown in
Fig.~\ref{fig:models}(c), the dipole moment is given by
\begin{eqnarray}
p_{\rm C}/e &=&
 -5\alpha + 4\beta +6(\alpha-\delta_{X})-6(\beta-\delta_{\textrm{TE}})
 \nonumber\\
 &=&
 \alpha - 2\beta -6(\delta_{X}-\delta_{\textrm{Te}}),
\label{eq:ModelC}
\end{eqnarray}
and comparing with Eq.~(\ref{eq:ModelA}), this is just
\begin{equation}
p_{\rm C} = p_{\rm A} +6e(\delta_{\textrm{Te}}-\delta_{X}).
\label{eq:AtoC}
\end{equation}

Turning to the results given in Table~\ref{tab:2model}, we see
that this analysis agrees well with the
full DFT results.
We can now think of Model A as a limit in
which $\delta_{X}\!=\!\delta_{\textrm{Te}}\!=\!0$, and Model B
as corresponding to $\delta_{X}\!=\!0$ and
$\delta_{\textrm{Te}}\!=\!5\beta/12$.  Clearly, neither is a
good approximation.  From another point of view, we can say that
an accurate picture of the dipole is given by modifying Model
A according to Eq.~(\ref{eq:AtoC}). Note, however, that $\delta_{\textrm{Te}}$
is much larger than $\delta_{X}$ in Table~\ref{tab:2model},
as expected given the stronger covalency of the Bi-Te bonding,
as was discussed toward the end of Sec.~\ref{sec:struct}
when describing the Wannier functions
shown in Fig.~\ref{fig:struct}(c).
Thus, Eq.~(\ref{eq:AtoC}) leads to a large positive correction
to the prediction of Model A.


\subsection{Piezoelectric response at fixed in-plane lattice constant}

\begin{figure*}
	\centering
	\includegraphics[width=17.8cm]{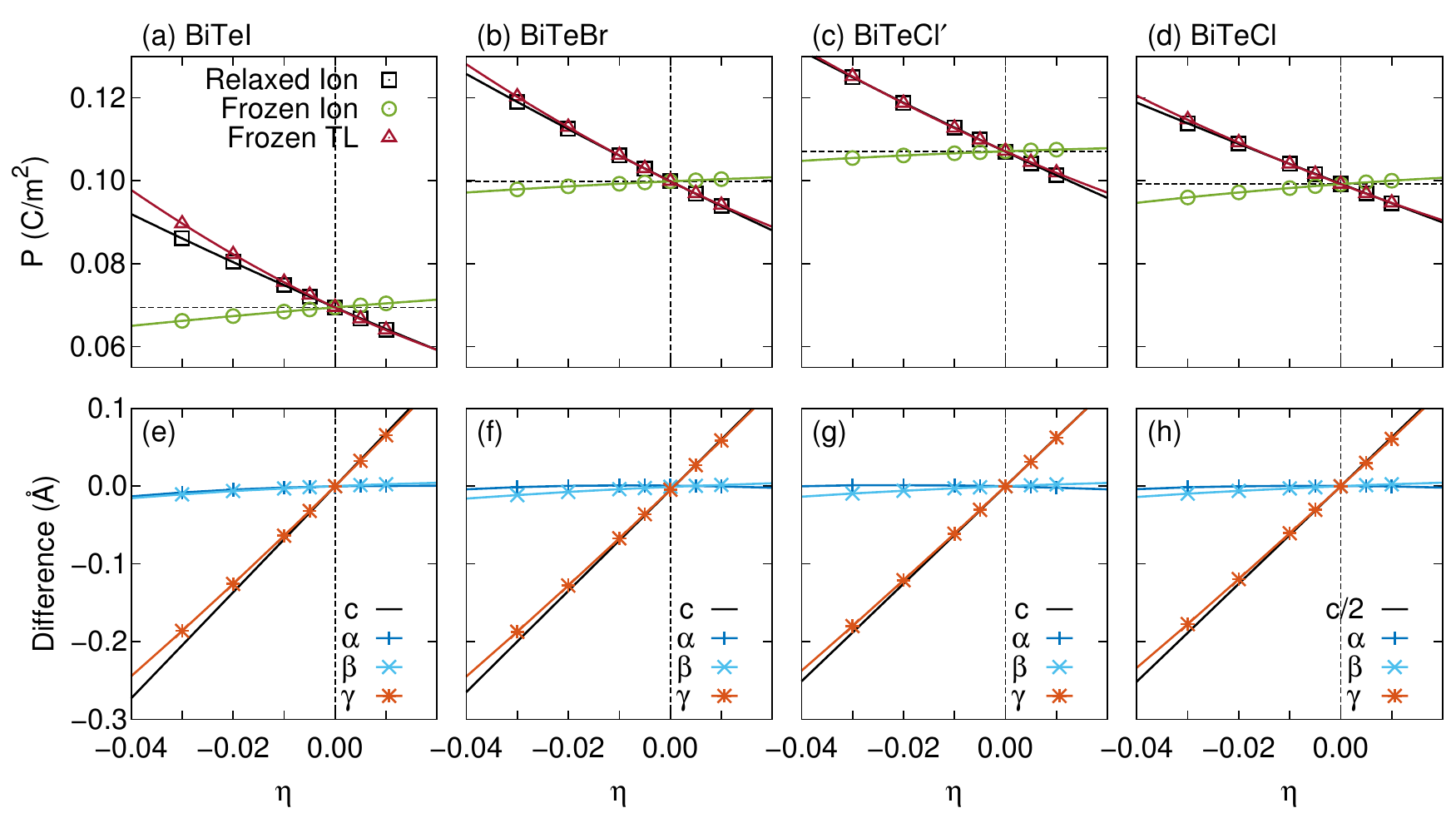}
	\caption{(a-d) Calculated polarization vs.\ strain
        $\eta=c/c_0-1$ of BiTe$X$ at
	fixed in-plane lattice constant for relaxed-ion,
	frozen-ion, and frozen-triple-layer cases.
        Solid lines are fits whose slopes
	at equilibrium ($\eta\!=\!0$, dashed vertical line)
	correspond to $e_{33}$.
	(e-h) Changes of relaxed structural
	parameters under uniaxial strain. Here $\alpha$ and $\beta$ are
        the Bi-$X$ and Bi-Te internal atomic interlayer distances
        (see Fig.~\ref{fig:struct}); $\gamma$ is
	inter-triple-layer spacing.}
	\label{fig:fixed_in_plane}
\end{figure*}

\begin{table}[t]
	\caption{Calculated piezoelectric responses of BiTe$X$
		materials: $e_{33}$ for relaxed-ion (RI), frozen-ion (FI),
		and frozen-triple-layer (FTL) cases; and proper, improper,
		and epitaxial $d_{33}$ piezoelectric responses. The 
		in-plane-lattice constants are relaxed for proper and
		improper $d_{33}$ cases and fixed otherwise.
	BiTeCl$'$ refers to the same $P3m1$ structure as for $X$\,=\,I
	and Br, although it is not the observed ground state for $X$\,=\,Cl.
    }
	\begin{ruledtabular}
	\begin{tabular}{ l c c c  c c c } 
		& \multicolumn{3}{c}{$e_{33}$ (C/m$^2$)} & \multicolumn{3}{c}{$d_{33}$ (pm/V)} \\
		& RI & FI & FTL & Prop. & Imp. & Epi. \\ [0.5ex]
		\hline
		BiTeI  & $-$0.53 & 0.10 & $-$0.57 & $-$24.4 & $-$23.0 & $-$24.4 \\
		BiTeBr & $-$0.61 & 0.06 & $-$0.60 & $-$30.6 & $-$28.6 & $-$31.2 \\
		BiTeCl$'$& $-$0.57 & 0.04 & $-$0.54 & $-$34.6 & $-$32.4 & $-$35.9 \\
		BiTeCl & $-$0.47 & 0.09 & $-$0.47 & $-$27.6 & $-$25.1 & $-$29.7
	\end{tabular}
	\end{ruledtabular}
	\label{tab:piezos}
\end{table}

\subsubsection{Piezoelectric stress coefficients}

Figures~\ref{fig:fixed_in_plane}(a-d) show the 
polarization of BiTe$X$ as a function of
uniaxial strain at fixed in-plane lattice constant.
These values are labeled ``relaxed-ion" since they
include full structural relaxation.  
The slopes at the equilibrium
state correspond to the piezoelectric stress tensor elements $e_{33}$,
which are reported as the relaxed-ion (RI)
values in the first column of Table~\ref{tab:piezos}.
The values are strongly negative for all three materials.

In the theory of piezoelectricity, it is a common
practice to decompose the response into a ``frozen-ion'' component,
defined by uniformly scaling the atomic positions in the unit cell
as the strain state is changed, and a ``lattice'' contribution that
comes from the change of internal atomic coordinates with strain.
The polarizations predicted by a naive
frozen-ion model (green circles) in  Figs.~\ref{fig:fixed_in_plane}(a-d)
are far from correct, and the corresponding $e_{33}$ values shown
in Table~\ref{tab:piezos} are much too small, and have the wrong sign,
compared to the full relaxed-ion results.

The above decomposition is in fact not appropriate for these quasi-2D systems.
To see why, we first look at the changes of
the ionic coordinates shown in Figs.~\ref{fig:fixed_in_plane}(e-h).
It is clear that the Bi-$X$ and Bi-Te interlayer spacings ($\alpha$ and
$\beta$) are almost constant as a function of $\eta$, while the
changes in the inter-TL spacing $\gamma$ tracks very closely with
the $c$ lattice constant.  This shows that a frozen-TL model,
in which the internal coordinates of the TL are fixed while only
the spacing between them changes, is an excellent approximation for
these systems.
The accuracy of this model can be understood as arising from the weak
van der Waals bonding between TLs, in contrast to the much
stronger covalent bonding within the TLs.
The polarizations obtained from the frozen-TL model are shown
as the triangles in Figs.~\ref{fig:fixed_in_plane}(a-d), and the
$e_{33}$ values are given in the `FTL' column of Table~\ref{tab:piezos}.
These are in excellent agreement with the full
first-principles (`RI') results, both in sign and magnitude.

\comment{\Green{It is important to understand, however, that this
frozen-TL model does not entail a freezing of the internal electronic
charge distribution within the TL.
That is, we allowed electronic relaxation by recomputing the electronic
structure self-consistently at each value of $c$ when presenting
the results shown above.  An important question, then, is whether
the dipole of the TL remains approximately constant as a function of
the uniaxial strain, or whether it changes significantly.}}

It is important to note that the frozen-TL model does not imply
a fixed dipole for the TL, as electronic relaxation occurs self-consistently at each value of $c$.
If the dipole of the frozen TL were independent of $c$, then
the negative piezoelectric response would be described as
coming from a simple volume effect, where the change in
$P=p/V$ is mainly a result of the change in $V$, with the
result that a compression or expansion simply concentrates
or dilutes the polarization.  More generally, the change in
dipole can be taken into account by writing
\begin{eqnarray}
e_{33}&=&\frac{\partial}{\partial\eta}\Big(V^{-1}p\Big)_{\eta=0}\nonumber\\
&=&-\frac{p_0}{V_0}+\frac{1}{V_0}\,\frac{\partial p}{\partial\eta}\Big|_{\eta=0}
\label{eq:decomp}
\end{eqnarray}
where $p_0$ and $V_0$ are the dipole moment and cell volume at
$\eta=0$.
If the polarization $p$ of the frozen TL were independent of
$\eta$, then only the first term would be present.  This purely
mechanical model, based only on the change of volume, does
correctly predict the negative sign of the piezoelectric response,
but we find that it severely underestimates the magnitude of the
effect.

To investigate this, we have calculated the dependence of
the Wannier-center shifts $\delta_{\textrm{Te}}$ and $\delta_X$ on
the uniaxial strain $\eta$,
since these are the parameters that reflect the internal
change of the TL dipole that is not captured by the structural
coordinates alone.
In the purely mechanical limit, $\delta_{\textrm{Te}}$ and $\delta_X$
would be independent of $\eta$,
and the second term in Eq.~(\ref{eq:decomp}) could be dropped.
We find that $\delta_{\textrm{Te}}$ is indeed very nearly
independent of $\eta$; it
varies by only about 0.1\% with a 1\% change in $c$. By contrast, we
find a much more significant change in the position of the
centers of the Bi-$X$ Wannier functions, with $\delta_X$ changing by
about 2.3, 2.6, and 2.1\% for $X$=I, Br, and Cl respectively,
for every 1\% change in $c$.  We can rationalize this change by
noting that the halogen environment becomes more symmetric
(less distinction between intra-TL and inter-TL neighbor
distances), so that the Wannier center shifts toward the $X$
coordinate, as the TLs are pressed closer to each other.
An inspection of the changes of the Wannier functions (not shown)
does indicate a stronger $X$-Te hybridization across the van der
Waals gap, consistent with a reduction of $\delta_X$, as the distance
between TLs is decreased.

In short, we find that there is a substantial increase in the
magnitude of the slab dipole with compression, which comes about
because of the change in $\delta_X$ values as the TLs are pressed
together. The consequences of this are summarized in
Table~\ref{tab:dipole}, where the contributions of the first
and second terms in Eq.~(\ref{eq:decomp}) are given independently,
clarifying their relative contribution to the piezoelectric
response $e_{33}$.  It now becomes clear that while the purely
mechanical model, represented by $-p_0/V_0$, correctly gives
the negative sign of the piezoelectric response, it underestimates
the magnitude of the response by almost an order
of magnitude. The modulation of the dipolar contribution of
the Bi-$X$ bond with compression provides a large boost to the
observed effect, and is responsible for the unexpectedly large
piezoelectric effect that is revealed by our calculations.

\subsubsection{Piezoelectric strain coefficients}

Next, we consider the piezoelectric strain response $d_{33}$.
Because this is defined under zero-stress boundary conditions,
the unit cell area now varies with $\sigma$, and there is
a distinction between proper and improper
piezoelectric responses~\cite{Vanderbilt2000}.
The ``improper'' piezoelectric response 
$d_{33}^{\textrm{imp}} = {\partial P}$/${\partial \sigma}$
simply describes the change of polarization under stress. 
However, the piezoelectric response is typically measured by
tracking the stress-induced current flowing between top and
bottom electrodes in a capacitor configuration, which
corresponds to the change of surface charge per unit cell
$\tilde{P} = AP$ with external stress, where $A$ is the cell
area.  This defines the
``proper'' piezoelectric response, where the two are related by
\begin{equation}
    d_{33}^{\textrm{prop}} = \frac{1}{A}\frac{\partial \tilde{P}}{\partial \sigma} = \frac{P}{A}\frac{\partial A}{\partial \sigma} + d_{33}^{\textrm{imp}}.
\label{eq:imp2prop}
\end{equation}

\begin{table}[b]
	\caption{Decomposition of the piezoelectric response into
		two contributions as given in Eq.~(\ref{eq:decomp}).
		All quantities in units of C/m$^2$.}
	\begin{ruledtabular}
		\begin{tabular}{ l c c c } 
			& $-p_0 / V_0$ & $V_{0}^{-1}(\partial p/\partial \eta)$ 
			& $e_{33}$ \\ [0.5ex]
			\hline
			BiTeI    & $-$0.069 & $-$0.460 & $-$0.53 \\
			BiTeBr   & $-$0.100 & $-$0.507 & $-$0.61 \\
			BiTeCl$'$& $-$0.107 & $-$0.464 & $-$0.57 \\
			BiTeCl   & $-$0.099 & $-$0.374 & $-$0.47 
		\end{tabular}
	\end{ruledtabular}
	\label{tab:dipole}
\end{table}

\begin{figure} [t]
    \centering
        \includegraphics[width=8.0cm]{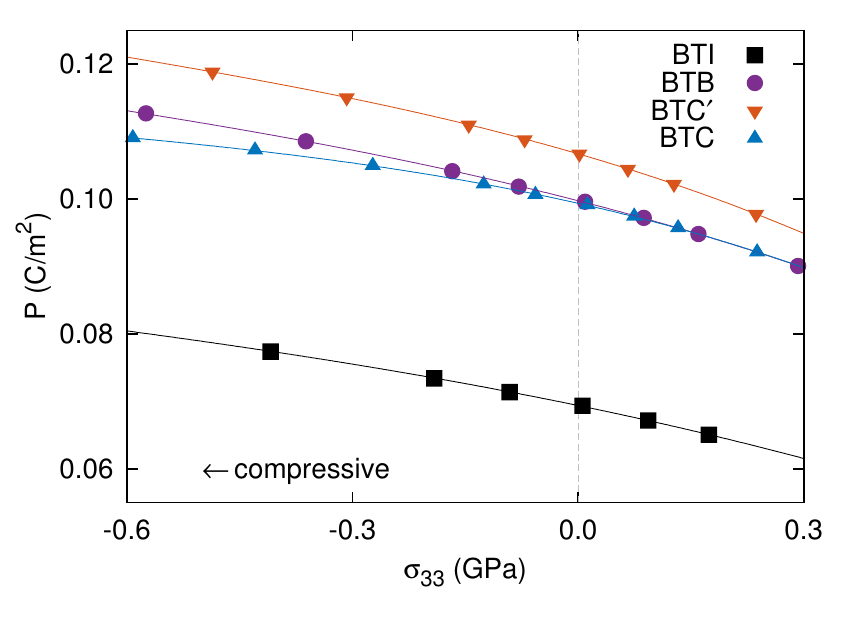}
        \caption{Calculated polarization of BiTe$X$
	vs.\ uniaxial stress at relaxed in-plane 
        lattice constant.
        Solid lines are fits whose slopes at equilibrium
        ($\sigma_{33} = 0$, dashed vertical line)
        correspond to $d_{33}$.}
    \label{fig:stress}
\end{figure}

Figure~\ref{fig:stress} shows the variation of $P$ with respect to
the uniaxial stress $\sigma_{3}$; the slope at $\sigma_{3} = 0$
yields $d_{33}^{\rm imp}$.  
A corresponding analysis
of the dependence of $\tilde{P}$ on uniaxial stress gives
$d_{33}^{\rm prop}$, and the results are
summarized in Table~\ref{tab:piezos}.
%
The improper response ranges from $-$23\,pm/V to $-$32\,pm/V,
which systematically increases in magnitude from $X$=I 
to Br to Cl$'$ in the same $P3m1$ structure. The change of structure
(Cl$'$ to Cl) gives a $\sim$20\% reduction, resulting in the 
Br compound having the largest $d_{33}^{\rm imp}$ response.
The correction term expressing the difference between $d_{33}^{\rm prop}$
and $d_{33}^{\rm imp}$ in Eq.~(\ref{eq:imp2prop}), whose sign
is negative (the cell area expands under uniaxial compression
along $c$),
enhances the negative piezoelectric responses slightly,
by $\sim$5-9\% compared to improper responses. As a result, 
the proper piezoelectric response of BiTeBr reaches $-$30\,pm/V.

\subsubsection{Epitaxial piezoelectric coefficients}

Finally, we calculated
the mixed response $d^\textrm{epi}$ (see the Methods section),
defined under conditions of fixed in-plane strain and out-of-plane
stress.  Here there is once again no distinction between proper and
improper responses.  The results are given in the last column of
Table~\ref{tab:piezos}.
The changes are not very dramatic; $d_{33}^{\rm epi}$ is similar
to $d_{33}^{\rm prop}$ for $X$\,=\,I \jkm{fixed a typo} and
slightly larger for the Br, Cl$'$, and Cl cases.
We have checked the effect of a fixed in-plane lattice constant that is set to a
modified value, as in the case of coherent heteroepitaxy on a substrate.
A compressive epitaxial strain is found to induce a negligible change,
whereas a tensile epitaxial strain reduces the magnitude of
the $d_{33}^{\rm epi}$ response somewhat, especially for $X$=Cl$'$ and Cl.

\subsection{Discussion}

To review, we have shown that the physics of the piezoelectric
response in the BiTe$X$ system is very different from that of
conventional ferroelectrics such as perovskite oxides.
In those systems, the proximity to a polar instability, the
associated soft polar modes, and the anomalously
large dynamical effective charges generate very strong piezoelectric
responses.  Here, instead, we start with a system that is far
from any structural phase transition, so it might be expected to
show quite a small piezoelectric response.

Nevertheless, we find a substantial piezoelectric response in
the BiTe$X$ system.  Our theory shows that
a model in which the internal structure
of each TL is frozen, and only the spacing between them
changes, gives an excellent account of the structural
changes under applied uniaxial strain.  If the dipole moment
of the TL were also frozen, this would already account for the
anomalous negative sign of the piezoelectric response. However,
we find that the electronic charge redistribution within the TL plays
a very important role, and is responsible for the
surprisingly large $e_{33}$ values.

To be sure, our calculated $|e_{33}|$ values of $\sim$0.50\,C/m$^2$
are smaller, by an order of magnitude or more, than those of well-known
perovskites such as PbTiO$_3$ and PZT, which have values
in the range of 4-12\,C/m$^2$.
However, it is still comparable to that of
wurtzite semiconductors (0.02-1.5\,C/m$^2$) and $ABC$ 
ferroelectric materials (0.4-1.5\,C/m$^2$).~\cite{Bernardini,Liu2017}
AlN and LiMgAs are reported to have the largest $e_{33}$ responses
of $\sim$1.5\,C/m$^2$ among each class of 
materials,\cite{Bernardini,Liu2017}
which is slightly larger than the classical wurtzite piezoelectric
material ZnO having $e_{33}\simeq$\,1.2\,C/m$^2$,\cite{Wu2005,Catti2003}
and only a factor of three larger than that of BiTe$X$.
%
%
The most negative $e_{33}$ among the $ABC$ ferroelectrics is
found in NaZnSb as $e_{33}$\,=\,$-$1.04\,C/m$^2$, only twice
larger than BiTe$X$.
A theoretical investigation of the negative piezoelectric responses
in the $ABC$ ferroelectrics has revealed that the frozen-ion
$\overline{e}_{33}$ gives a negative contribution that dominates
the total response.~\cite{Liu2017}  This is
in a sharp contrast with BiTe$X$, where the frozen-ion 
contribution is rather small and positive. 

%
Moreover, the piezoelectric response of BiTe$X$
is further magnified when converting to the
piezoelectric strain coefficient $d_{33}$ because of the softness
of the interlayer van der Waals interaction, which implies a
large strain per applied uniaxial stress.  This makes the comparison
of the $d_{33}$ values of the BiTe$X$ materials even closer to being
competitive with other piezoelectrics.
Our calculated $|d_{33}|$ values, in the range of
24-36\,pm/V, compare well to those of LiMgP (25\,pm/V) and
LiMgAs (29\,pm/V), which have the most positive $d_{33}$
among the $ABC$ ferroelectric materials,
even though their $e_{33}$ values are a factor of three larger as
discussed above.~\cite{Liu2017}
The relative enhancement of $d_{33}$ relative to $e_{33}$ in
BiTe$X$ is also evident by comparison with BaTiO$_3$,~\cite{Wu2005}
where $|e_{33}| = 4.44$\,C/m$^2$ is an order
of magnitude larger than for BiTe$X$, while $|d_{33}| = 14.7$\,pm/V
is a factor of two smaller than for BiTe$X$.

Recent studies have shown that ultrahigh piezoelectric responses have
been achieved with $d_{33}$ up to 2,800\,pm/V in PMN-PT~\cite{Park1997,Zhou2014}
and 640\,pm/V in BaTiO$_3$-based ceramic systems,~\cite{Praveen2015,Wei2018}
and a negative piezoelectric response of $d_{33}\!=\!-$690\,pm/V
has been reported in the classe of ferroelectric polymers based
on polyvinylidene difluoride (PVDF).~\cite{Ghosh2015,Wei2018}
We note, however, that these ultrahigh piezoelectric responses are
a result of careful compositional tuning of the system.
For example, the simple $\beta$ phase of PVDF exhibits a $d_{33}$
of $-$50\,pm/V, which is only a factor of two larger than
for BiTe$X$.~\cite{Soin2015,Wei2018}

We also note that the $d_{33}$ responses of BiTe$X$
are comparable to the values reported 
for some thin-film PZT samples (21.3\,pm/V)~\cite{Guo2013}, 
although still an order of magnitude
smaller than for thick-film PZT (457\,pm/V).
Because the mechanism of piezoelectricity in the BiTe$X$ system
is completely independent of any soft-mode transition, there is no
reason to expect our computed responses to suffer from the kind of
finite-size effects that suppress the piezoelectricity in thin-film
geometries for conventional materials.  Taken together with our
encouraging estimates of the size of the responses reported above,
these results suggest that the BiTe$X$ materials could
provide a promising alternative material system to use as a
basis for microactuator and other specialized applications.

\section{Summary}

We have used first-principles methods to calculate the
polarizations and piezoelectric responses of BiTe$X$ for $X$=Cl,
Br, I.  The piezoelectric response is found to be negative.
The change in structure under uniaxial stress is described to an
excellent approximation by a ``frozen triple-layer''
model in which the BiTe$X$ unit is internally rigid,
while the spacing between these units is modulated by the
applied uniaxial strain. However, the dipole moment of the TL is
not frozen, and changes with stress due to electronic relaxation
are found to dominate the piezoelectric response.
The piezoelectric responses are an order of magnitude
smaller than those of commercial bulk piezoelectric materials,
but the mechanism can be expected to survive in the thin-film
limit where standard piezoelectrics tend to degrade.  Thus,
BiTe$X$ could be a promising alternative material for thin-film
piezoelectric devices. The recent explosion of interest in stacked
van-der-Waals--bonded heterostructures provides opportunities for
BiTe$X$ as a piezoelectric component in such systems.

\begin{acknowledgments}
This work is supported by the
ONR Grants N00014-16-1-2951 and N00014-17-1-2770.
\end{acknowledgments}

%

\end{document}